\def\ba{\begin{eqnarray}}
\def\ea{\end{eqnarray}}
\def\lb{\label}
\def\nn{\nonumber \\}
\def\bi{\bibitem}

\def\g{\gamma}

\def\rr{\rightarrow}

\def\pt{p_\perp}

\def\pr{\perp}
\def\sp{\;\;\;\;}

\documentclass[12pt]{article}
\usepackage{graphicx}
\begin{document}
\title{Tsallis $p_\pr$ distribution from statistical clusters}

\author{A.Bialas
\\ M.Smoluchowski Institute of Physics\\Jagellonian University\thanks{Address: Lojasiewicza 11, 30-348 Krakow, Poland; e-mails: bialas@th.if.uj.edu.pl}
}
\maketitle



\begin{abstract}

It is shown that  the   transverse momentum  distributions of particles emerging from the decay of  statistical clusters, distributed according to a power law in their transverse energy, closely resembles that following from the Tsallis non-extensive statistical model.  The experimental data are well reproduced with the cluster temperature $T\approx 160$ MeV.
\end{abstract}

{\bf 1.} It is now well-documented that the transverse momentum distribution of various particles at high energy and in a very broad range of the transverse momentum,  is correctly described by the Tsallis distribution. This was observed  in all high-energy experiments \cite{phenix1}-\cite{cms1}, as well as in the recent phenomenological analyses \cite{cl1}-\cite{rw}.   This observation is usually interpreted in terms of the statistical model of particle production, employing the Tsallis non-extensive statistics \cite{tsallis,w2}.  Such interpretation\footnote{The fundamental relation between the Boltzmann and Tsallis statistical models is clearly explained in  \cite{w1}.}, although very attractive, meets a serious difficulty, however: it is indeed not easy to explain why  any statistical model can apply at very large transverse momenta, where the perturbative  QCD phenomena are known to dominate.

This  problem was recently addressed in a series of papers \cite{ww1,ctww2} where the ideas derived from perturbative  QCD (as applied to hard interactions), accompanied by the parton cascade (responsible for the jet fragmentation)  were used to explain this puzzling result. 

In this note, following  the general  idea suggested in \cite{ww1,ctww2} (c.f. also \cite{w4}), we apply it to the statistical model of particle production which is  rather successful in describing  data  on particle multiplicities (For a review, see, e.g. \cite{becattini1,and}).   We   show that the Tsallis formula can be recovered, to a  good accuracy, in the model where the observed particles are  decay products of     clusters \cite{bclus, bp, becattini2} which (i) decay according to the standard Boltzmann statistics and (ii) the distribution of their Lorentz factors  follows a power law, as suggested by (perturbative) QCD. This observation indicates that the experimental validity of the Tsallis formula may be interpreted as another confirmation of the standard statistical model rather than that of its non-extensive Tsallis version\footnote{Another approach aiming at the explanation of the power law tails within statistical model is discussed in \cite{bgg}.}. 

Indeed, the intriguing "unreasonable" success of the statistical model in description of  multi-particle production in various processes and  at various energies    suggests that  the final stage of  the process of hadronization is  dominated by the  hadrons in  the state of statistical  equilibrium.  It is also clear that the equilibrium cannot be global, as the observed spectra are far from isotropic. These observations lead naturally to the idea \cite{bclus} that the transition from the early state of the process, dominated by interactions between  the hadronic constituents, most likely proceeds through an intermediate stage of  clusters  emitting the final hadrons according to the rules of statistical physics. 

If one admits that this process of cluster formation and thermal decay is a universal feature of hadronization, one is led to the conclusion that also the high transverse momentum jets  hadronize in the same way (c.f. \cite{bp}). It follows that the characteristic features of clustering should  leave their imprints even in the region  of hard physics. In the present paper we show that this picture, when combined with the power law distribution of the (transverse) Lorentz factor  of the cluster,  leads to transverse momentum distribution of the decay products which is very close to that of Tsallis (and thus also close to experiment). 

In the next section the idea of the statistical cluster is formulated and the transverse momentum distribution of its decay products is derived. The relation to the Tsallis distribution is discussed in Section 3.  Summary and comments are given in the last section.

{\bf 2.} Following the ideas explained above,  the decay distribution of the statistical cluster at rest is taken in the form of the Boltzmann distribution which, for a cluster  moving with the four-velocity $u^\mu$ becomes
\ba
\rho(p;u)d^2\pt dy= e^{-\beta p_\mu u^\mu }d^2\pt dy  \lb{single}  \lb{mc}
\ea
where $\beta=1/T$. 

Consider a cluster at rapidity $Y$ moving in the transverse direction with the velocity $v_\perp$. 
 We have
\ba
u_0=\sqrt{1+u_\pr^2}\cosh Y;\sp u_z=\sqrt{1+u_\pr^2}\sinh Y; \sp v_z=\tanh Y;\nn
u_\pr=\g v_\pr;\sp \g=(1-v^2)^{-1/2}\;\rr\; \sqrt{1+\g^2v^2}=\g.
\ea
 The distribution of particle momentum is then
\ba
\rho(p,y) dy= dy  d^2p  \int d^2v_\pr dY G(v_\pr, Y)e^{-\beta \g_\pr m_\pr \cosh(y-Y)-\beta p_\pr u_\pr\cos\phi}
\ea
where $\phi$ is the angle between $v$ and $p_\pr$ and where we have denoted 
\ba
\g_\pr\equiv \sqrt{1+u_\pr^2}=\g\sqrt{1-v_z^2}\;\rr\; u_\pr=\sqrt{\g_\pr^2-1}
\ea
 Integration over $\phi $ and $y$  gives
the distribution of the transverse momentum: 
\ba
\rho(p)   d^2p=    d^2p\int d^2v_\pr dY G(v_\pr, Y) K_0[\beta m_\pr\g_\pr]I_0[\beta p_\pr u_\pr]
\lb {klkt}
\ea

  \begin{figure}[h]
\begin{center}
\includegraphics[scale=1.0]{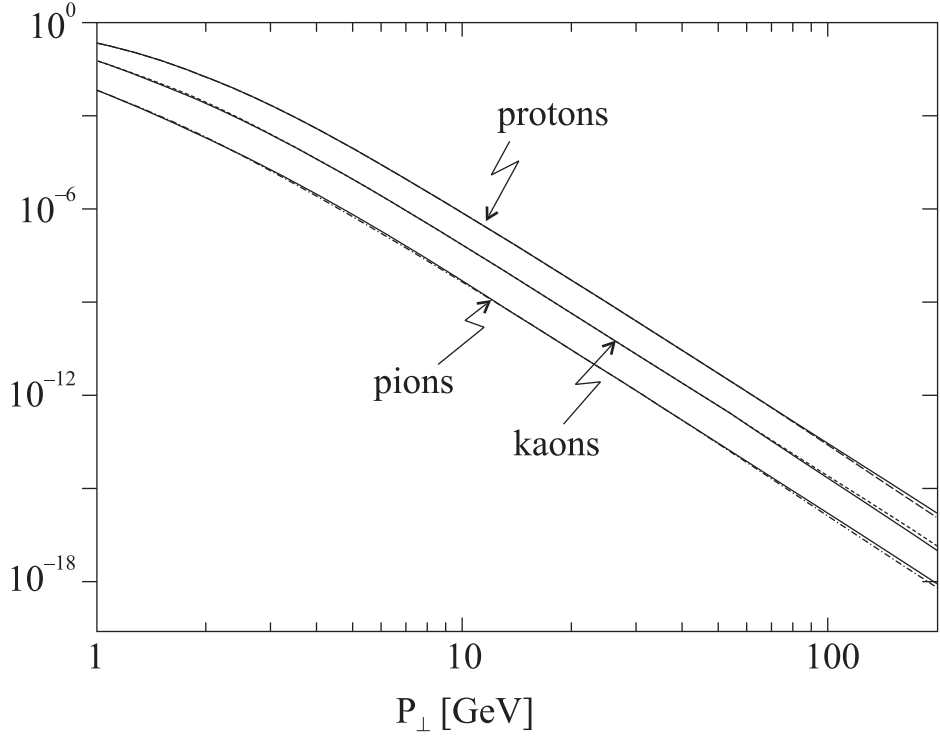}
\end{center}
\caption{Transverse momentum distribution of pions, kaons and protons from the statistical cluster decay (dashed lines), normalized to 1 at $p_\pr=0$, compared  to two  Tsallis distributions (Eq.(\ref{tsver})) (full lines).  1 GeV $\leq p_\pr\leq 200$ GeV. $T=155$  MeV, $\kappa$=6.5. Best fit from $p_\pr=0$ to $p_\pr=50$ GeV.}
\label{pikp200}%
\end{figure}

  \begin{figure}[h]
\begin{center}
\includegraphics[scale=1.0]{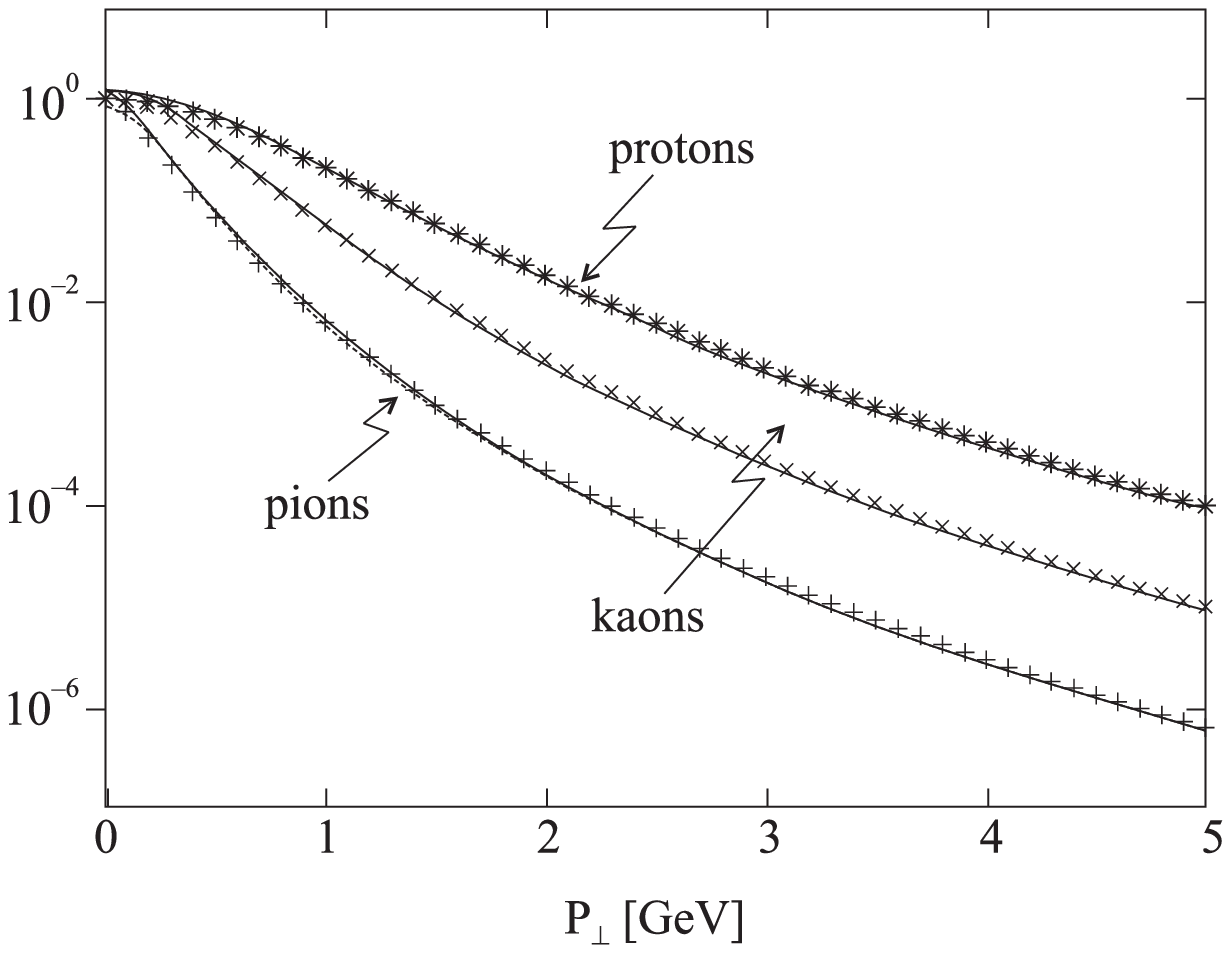}
\end{center}
\caption{Same as Fig.\ref{pikp200} but for $0\leq p_\pr \leq 5$ GeV. Lines: the Tsallis distribution.
Crosses and stars: statistical clusters. }
\label{pikp5}%
\end{figure}

{\bf 3.} To evaluate the  distribution of  transverse momenta of the cluster decay products, one needs the distribution of the cluster transverse  velocity $v_\perp$.  In this paper we study a power law in the transverse Lorentz factor $\g_\perp$ (for a fixed cluster mass, this would correspond to a power law in its transverse energy). Thus we take
\ba
d^2v_\pr dY G(v_\pr, Y)\sim G(Y) dY \g_\pr ^{-\kappa} d\g_\pr  \lb{gv}
\ea
Given simplicity of this  assumption, it was  rather surprising to find that it leads to the distribution which closely resembles that of  Tsallis\footnote{Qualitatively, the  result  of this kind may be actually expected,  as  it is   the well-known  \cite{w1,ww4,biro} that the Tsallis formula is  naturally obtained by adequate fluctuations of the  parameters of the Boltzmann spectrum.}, from $p_\perp \approx 100$ MeV up to  $p_\perp$= 200 GeV. 
This was verified numerically for the cluster temperature in the region from 100 till 180 MeV and the power $\kappa$ from 4 till 7, i.e. in the range covering  the physical conditions one may expect in  high-energy collisions..

An example of such calculation is shown  in Figs \ref{pikp200} and \ref{pikp5} where the distributions of pions, kaons and protons  evaluated using (\ref{klkt}) and (\ref{gv}) with $\kappa=6.5$ and T=155 MeV, are compared with the two versions of the Tsallis distribution  \cite{cl2,cl4,w1,ww4}:
\ba
D_1= c m_\perp[1+(q-1)m_\perp/T_{ts}]^{q/(1-q)};\nn
D_2= c[1+(q-1)m_\perp/T_{ts}]^{1/(1-q)},   \lb{tsver}
\ea
where $c$ is the normalization constant, $q-1$ measures the deviation from the standard statistical model and $T_{ts}$ is the Tsallis temperature\footnote{The form $D_1$ is obtained by demanding   maximum of the Tsallis entropy, i.e. thermodynamic equilibrium \cite{mmd,pp}. The second form is  the standard Tsallis distribution.}.

One sees that, except at very small $p_\perp$, below $\sim$ 100 MeV,  there is an  excellent agreement between the two formulations and for all kinds of particles. One also sees that for $p_\perp \geq 1$ GeV it is difficult to distinguish between the two versions of the Tsallis distributions. For  the distribution $D_1$ the Tsallis parameter $T_{ts}$ can be approximated by the simple relation $T_{ts}\approx (q-1) T$. This is not true, however, for $D_2$. In this case   the relation between $T_{ts}$ and $T$ is more complicated and, moreover, it depends substantially on the particle mass.

\begin{figure}[h]
\begin{center}
\includegraphics[scale=1.0]{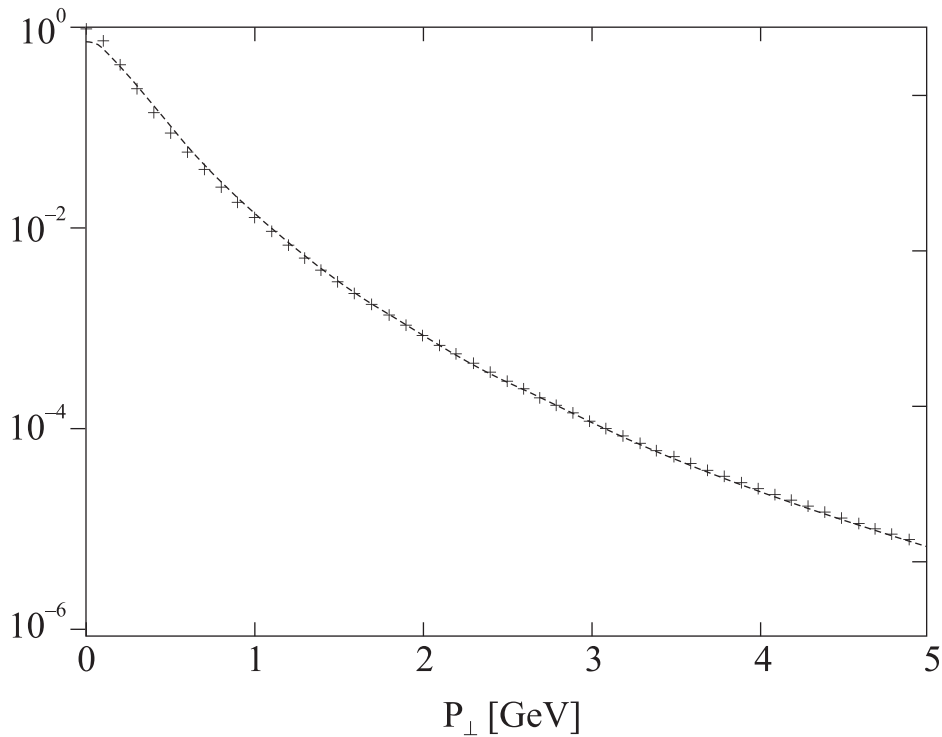}
\end{center}
\caption{Transverse momentum distribution of charged particles from the statistical cluster decay (crosses),  compared  to the  Tsallis distribution (dashed line) used in \cite{cl2} (the first formula in (\ref{tsver})).  0 $\leq p_\pr\leq 5$ GeV. $T=155$  MeV, $\kappa$=6.5. Best fit from $p_\pr=0$ to $p_\pr=50$ GeV.}
\label{t5}%
\end{figure}

Recently, a new analysis of transverse momentum distribution of charged particles  in terms of the Tsallis distribution has been published \cite{cl2}. 
To compare these results with our approach, we have  evaluated 
the distribution following from the decay of a cluster for pions, kaons and protons and constructed the distribution of charged particles, using the weights (1:1:2), as proposed in \cite{cl2}. In Fig. \ref{t5} the results in the region from $p_\perp=0$ till $p_\perp=5$ GeV are compared with the Tsallis distribution from \cite{cl2}. One sees that the agreement is very good, except at $p_\perp < $ 100 MeV.  The parameters of the Tsallis distribution in this case are $q-1=0.150$ and $T_{ts}$=76 MeV, in good agreement with \cite{cl2}. The region $p_\perp \geq 5$ GeV is not shown because in this region one simply cannot distinguish between the two curves.

{\bf 4.} In summary, we have discussed the  transverse momentum distributions of particles emitted in the decay of  a statistical cluster. It was shown that if the (transverse) Lorentz factor of the cluster follows a power law,  the resulting distribution is very close to that derived from the Tsallis non-extensive statistics. 

This result may be considered as a possible  explanation of  the surprising observation that the Tsallis formula  works not only at small transverse momenta (where the ideas of statistical equilibrium may be applicable) but even at transverse momenta as large as $\sim$ 200 GeV.  

Some comments are in order.

(i)  It should be emphasized that the observed similarity between the Tsallis formula and that following from the statistical cluster decay, is only an approximation. Our results indicate, however,  that it may be rather difficult to distinguish experimentally between these two approaches. Perhaps the measurements at larger transverse momenta may be helpful, as the two distributions start to deviate  from each other at energies above 200 GeV.

(ii)  We have been discussing emission of a single statistical cluster. 
 As it is rather unlikely that a high-energy jet may fragment into a single cluster, production of many clusters  must also be considered.  Since our  discussion concerns only the single-particle distribution,  however, the results are  insensitive to the number of clusters produced in a given event, provided they are emitted independently.

(iii) Clearly, the power law assumed in  (\ref{gv}) is only a phenomenological guess  and should be treated as such. Its main advantage is the extreme simplicity  (for  more elaborate calculations see, e.g., \cite{ww1,ctww2,w4}). Needless to say, the parameter $\kappa$ remains free  at the present stage, and cannot be reliably   evaluated from theory. 

(iv) It has been shown recently  \cite{mclpr1} that the distributions of transverse momenta  at various energies follow a scaling law, suggested by the saturation property of the parton distributions.   An interpretation of this  observation in terms of the Tsallis approach was proposed in \cite{w4,w3}. It would  be thus interesting to investigate how this scaling property of the spectra  translates into the results shown in the present  paper.

ACKNOWLEDGEMENTS

The invaluable guidance of Grzegorz Wilk through the intricacies of the Tsallis approach to particle production and through the abundant literature on this subject is highly appreciated. 
I also thank Francesco Becattini, Adam Bzdak, Marek Gazdzicki  and Robi Peschanski for very helpful  discussions and encouragement.
This investigation was supported in part by the Polish National Scientific Center  (Narodowe Centrum Nauki),  DEC-2013/09/B/ST2/00497.


\end{document}